\documentstyle[pre,floats,aps]{revtex}
\input{psfig} 
\begin{document}
 
\def\B{\mbox{\bf B}}
\def\x{\mbox{\bf x}}
\def\z{\mbox{\bf z}}
\def\u{\mbox{\bf u}}
\def\j{\mbox{\bf j}}
\def\A{\mbox{\bf A}}
\def\e{\bf {\hat e}}
\def\E{\bf {\hat E}}
\def\m{\bf {\hat m}}
\def\M{\bf {\hat M}}
\def\s{\bf {\hat s}}
\def\S{\bf {\hat S}}
\def\v{\bf v}
\def\zhat{\hat {\bf z}}
\def\psibar{\hat \psi}
\def\V{\mbox{\bf V}}
\def\u{\mbox{\bf u}}
 
\draft

\title{Design criteria of a chemical reactor based on a chaotic flow}
\author{X. Z. Tang\footnote{Email: tang@chaos.ap.columbia.edu}}
\address{Department of Applied Physics, Columbia University,
New York, NY 10027}

\author{A. H. Boozer}
\address{Department of Applied Physics, Columbia University, New York,
NY 10027\\
and Max-Planck Institut f{\"u}r Plasmaphysik, Garching, Germany}
\date{\today}

\maketitle

\begin{abstract}
We consider the design criteria of a chemical mixing device based on a 
chaotic flow, with an emphasis on the steady-state devices. 
The merit of a reactor, defined as the $Q$-factor, is related to the 
physical dimension of the device and the molecular diffusivity of the reactants
through the local Lyapunov exponents of the flow.
The local Lyapunov exponent can be calculated for any given flow field
and it can also be measured in experimental situations.
Easy-to-compute formulae are provided to estimate the $Q$-factor given either
the exact spatial dependence of the local Lyapunov exponent or its 
probability distribution function.
The requirements for optimization are made precise in the context
of local Lyapunov exponents. 
\end{abstract}


\section{Introduction}

A major concern in chemical engineering is to react two or more
species to form a new chemical product \cite{rosner_book}.
To increase production, one could boost the reaction rate
by a catalyst which is an optimization on a microscopic level.
One could also carry the reactants by a flow and maximize the 
diffusive effect that brings the chemicals together to react.
Traditionally the flow is designed to be turbulent to achieve 
higher mixing effect. Started in the 1980s, non-turbulent
but chaotic flows, are advocated to achieve the same effect with
significantly less energy consumption in driving the flow 
\cite{aref_84,ottino_book}. 
Much work has been done in this area
\cite{ottino_review,brenner_96,muzzio_96,chang_96}, 
but some of the fundamental
issues on optimizing the design  
still require a clarification. This paper intends to report
some progresses along this line. The emphasis on the local Lyapunov
exponent of the flow and its role in {\it quantitatively} determining 
the physics of both advective `mixing' and diffusive transport, 
is an unique aspect of this investigation.  

The typical reaction is a three species event, chemical $A$ 
reacts with chemical $B$ to form chemical $C,$
$$
A + B \rightarrow C.
$$
The quantities of interest are the concentrations or number densities
of each species, $C_A, C_B,$ and $C_C.$
The concentration of the end product is not independent and
can be expressed in terms of the history of $C_A$ and $C_B.$
If the chemicals are carried by a flow, the mathematical model for
the mixing-reaction process is 
the advection-diffusion-reaction equation.
Written out explicitly for each species, they are
\begin{eqnarray}
{\partial C_A / \partial t} + {\v}\cdot{\nabla C_A} 
& = & \nabla\cdot(D\nabla C_A) - \kappa C_A C_B; \label{a-d-r-A}\\
{\partial C_B / \partial t} + {\v}\cdot{\nabla C_B} 
& = & \nabla\cdot(D\nabla C_B) - \kappa C_A C_B.  \label{a-d-r-B}
\end{eqnarray}
In the general case, the chemicals react according to
$
m A + n B \rightarrow l C.
$
The concentrations in equations (\ref{a-d-r-A},\ref{a-d-r-B})
should then be replaced by the scaled concentrations
$C_A' \equiv C_A/m, C_B' \equiv C_B/n, $ and $C_C'\equiv C_C/l.$

We will assume that the original carrier flow is sufficiently 
energetic that the back-reaction of the reaction process on the 
background flow is negligible. This assumption decouples
the third governing equation, for example, 
the Navier-Stokes equation for the flow field,
from the two coupled advection-diffusion-reaction 
equations (\ref{a-d-r-A},\ref{a-d-r-B}).
The principles for the design optimization, are then found
by solving the advection-diffusion-reaction equations and identifying 
the flow features that crucially affect the production rate and 
quality. 

The underlying physics is better explained by a transformation of the variables
of the original advection-diffusion-reaction equations.
Define $\phi \equiv C_A - C_B$ and $f \equiv C_A + C_B,$
one has
\begin{eqnarray}
{\partial\phi/\partial t} + {\v}\cdot\nabla\phi
& = & \nabla\cdot (D\nabla \phi), \label{a-d-equation} \\
{\partial f/\partial t} + {\v}\cdot\nabla f
& = & \nabla\cdot (D\nabla f) - {1\over 2} \kappa (f^2 - \phi^2).
\label{a-d-r-equation}
\end{eqnarray}
Obviously $f^2-\phi^2=4C_A C_B \ge 0$ and $f \ge \|\phi\|.$

There are generally two classes of reactors, which we will call a closed
flow system and an open flow system. A canonical closed flow system is
a stirred tank \cite{tatterson_94}.
Mathematically it corresponds to an initial value problem, with simple
boundary conditions. The degree of mixing is measured over time.
A canonical open flow system is a tubular device such as 
an automobile catalytic converter. 
Raw exhaust gas
constantly flows in, the poisonous elements are removed by chemical reactions, 
and the treated gas is discharged to the tail pipe.
The system is expected to function in steady state. Mathematically
it is a time-independent boundary value problem.
The integral form of equations (\ref{a-d-equation},\ref{a-d-r-equation})
makes the global balance transparent.
For an engineering device operating in steady state,
$$
\int_V {\partial f\over{\partial t}} d^3{\x}
= {\partial\over{\partial t}} \int_V f d^3{\x} = 0.
$$
The remaining part is
$$
\int [f {\v} - D\nabla f]\cdot d{\bf a} = - {1\over 2}\kappa
\int_V (f^2 - \phi^2) d^3{\x}.
$$
Here we have assumed incompressibility $\nabla\cdot{\v}=0$ for
mathematical simplicity.
The left hand side is the rate of new reactant influx and
the right hand side is the rate of reactant depletion due to
reaction.
Naturally for a steady state device they balance each other out.

The fast reaction scenario would further reduce the 
mathematical complexity and help clarifying the basic issues.
If the chemical reaction rate is sufficiently fast that
\begin{equation}
\label{fast_reaction}
\kappa \gg - {\int [f{\v}-D\nabla f]\cdot d{\bf a} \over
	{\int_V (f^2/2) d^3{\x}}},
\end{equation}
then $f \approx \|\phi\|$ everywhere in the system since
$$
\int_V [f^2 - \phi^2]d^3{\x} \ll \int_V f^2 d^3{\x}.
$$
Mathematically that is to say that the reaction
is fast so once $\phi$ is found, the problem is solved by
setting $\|\phi\| = f.$ 
Physically it says that two different reactants react
so fast that they can not coexist at the same point
anywhere in the device. Hence at least one of reactant
densities has to vanish locally and $\|\phi({\x})\|=f({\x})=Max(C_A,C_B).$
For engineering design purposes, 
the equivalent statement is that 
the reaction is sufficiently fast so that the efficiency of
the reactor is determined by the efficiency of mixing.
This is precisely the motivation for embedding the reactants
into a turbulent flow and more recently a chaotic one.
We note that the use of a catalyst is intended to boost
the microscopic reaction rate $\kappa.$

As expected, the inequality in equation (\ref{fast_reaction})
is equivalent to the statement that
the Damk{\"o}hler number is much greater than the Peclet number.
The Peclet number $Pe$ is the ratio between the characteristic diffusion
time $L^2/D$ and the advective time scale $L/U.$
The Damk{\"o}hler number $Da$ is the ratio between the characteristic
diffusion time and the typical reaction time $1/\kappa \langle f/2 \rangle$
with $\langle\cdots\rangle$ an average over space.
A reaction is fast if the typical reaction time is much shorter
than the advective time scale, {\it i.e.} $Da/Pe \gg 1.$
Although in a closed flow system like a stirred tank the ratio 
$Da/Pe$ might be 
less than one, it has to be  greater, and usually much greater, than one 
in a working steady-state device. This is a trivial statement of the
fact that the advective time scale $L/U$ in a steady-state device is simply
the duration for the reactants to stay in the reactor before discharge. 
It has to be longer
than the reaction time. Otherwise even a perfectly mixed reactants at the inlet
will not form much product at the time of discharge. In practice,
the requirement for a `fast' reaction $Da/Pe\gg 1$ is not difficult to
accommodate. For example it can always be 
satisfied in a tubular device
by simply increasing the longitudinal length $L.$   
Indeed, the minimum design constraint of a steady-state chemical reactor is to
have the longitudinal Damk{\"o}hler number greater, or much greater, than
the longitudinal Peclet number. Here longitudinal refers to the 
streamwise direction.  

The degree of global mixing in a reactor is measured by
\begin{equation}
\label{global_mixing_index}
\sigma(t) \equiv {1\over 2} \int_V \varphi^2 d^3{\x}
\end{equation}
with
$$
\varphi \equiv \phi - \int \phi d^3{\x} / \int d^3{\x}
$$
The separation between reactants is always non-negative so 
$\sigma \ge 0.$
The governing equation for the global mixing index is
$$
{d\sigma\over {dt}} = {\cal F} - \int D (\nabla\varphi)^2 d^3{\x}.
$$
The flux of separated chemicals into the system is
$$
{\cal F} \equiv - \int [{\varphi^2\over 2} {\v} 
- D \nabla {\varphi^2\over 2}] \cdot d{\bf a}.
$$

Now we can distinguish two kinds of mixing devices, closed
and open, which correspond to an initial value problem and 
a time-independent boundary value problem respectively.
For a closed system,  
$$
{\cal F} = 0, \,\,\,\,\, d\sigma/dt = - \int D (\nabla\varphi)^2 d^3{\x}.
$$
Obviously
$$
\sigma \ge 0 \longrightarrow \sigma = 0 \,\,\,\,
{\rm for\,\, sufficiently \,\,long\,\, time.}
$$

An open system is markably different.
For example ${\cal F}$ normally is positive.
It must be positive for steady state device where
$d\sigma/dt=0.$
Pictorially speaking, separated chemicals are brought in, 
$\varphi^2$ non-zero where
${\v}\cdot d{\bf a} < 0.$
Mixed and reacted chemicals are discharged, $\varphi^2$ essentially
zero where ${\v}\cdot d{\bf a} > 0.$

The flow design is driven by the rate of
$\sigma \rightarrow 0$ which measures the quality of the reactor.
The desire for a background flow can be clearly illustrated
in a closed system, where the relaxation of
$\sigma$ is completely due to diffusion driven
by the gradients of $\varphi.$ 
Since 
$$
(\nabla\varphi)^2 = {\varphi^2\over l^2} 
$$
the relaxation time scale is $l^2/ D.$
If there is no convection, $l$ is of reactor size and the 
relaxation time scale $l^2/D$ is hopelessly long.

Towards a better reactor, the carrier flow
must be designed to drive gradients in $\varphi$ or $\phi,$ which 
is an ideal task for a chaotic flow.
In addition, to achieve better quality of the end product,
it is desired that the gradients are driven `uniformly.'
In the case of a chaotic flow, the size of the KAM islands
should be minimized, which is an obvious direction for optimization. 
There is another, more subtle, issue concerning
the non-uniformity of the chaoticity and the existence of a barrier
to diffusion even within the chaotic region.
This subtlety can not be resolved by the standard Poincare section
technigue, 
but it can be made mathematically precise in terms of
the finite time or local Lyapunov exponents of a flow. 
The goal of this paper will be to explain how to understand the merit
of a reactor by the
finite time or local Lyapunov exponent of the carrier flow.

Our approach based on the local Lyapunov exponents
should be contrasted with the past heavy reliance 
on the Poincare section as a diagnostic to measure the extent of mixing,
an approach that is prone to misinterpretation and does not provide any rate
information\cite{brenner_96}. 
The limitation of the Poincare section technique was 
explicitly noted by Swanson and Ottino in \cite{swanson}, and prominently
restated by Bryden and Brenner in \cite{brenner_96}. Muzzio and
Liu presented a numerical demonstration in a two dimensional
mixing-reaction problem\cite{muzzio_96}. 
The local Lyapunov exponent analysis, in our view, complements the
Poincare section by providing the exact rate information and removes
the ambiguities associated with the Poincare plots.     

The rest of the paper is organized as follows.
Section \ref{sec:finite_time_exponent} gives an introduction on 
how the finite time Lyapunov exponent enters the description of a
chaotic flow and the passive scalar transport.
Section \ref{sec:closed_flow} briefly describes the time-dependent
solution to the advection-diffusion equation, the more detailed analysis
can be found in \cite{tang} for the two dimensional case and 
\cite{tang_pof} for the three dimensional case.
The main body of the paper, which includes 
sections \ref{sec:steady_mixer}
\ref{sec:mixing_length} \ref{sec:local_exponent_Q}
\ref{sec:optimization_s_line} \ref{sec:num_example},
deals with the practically attractive steady-state reactors.
The main mathematical formulation has utilized cylindrical geometry and 
quasi-two-dimensional flows for clarity and convenience,
although the most important concepts and results can be readily
generalized.
Section \ref{sec:conclusion} highlights the main points of the
paper.

\section{Finite time Lyapunov exponent and a chaotic flow}
\label{sec:finite_time_exponent}

A chaotic flow is characterized by the exponential variations of 
neighboring streamlines. For example, the separation between 
neighboring fluid trajectories
of a divergence-free, time-periodic two dimensional flow obeys
\begin{equation}
\label{separation}
(dl)^2 = (d{\vec l}_0\cdot{\e})^2 \exp (2\lambda t)
	+ (d{\vec l}_0\cdot{\s})^2 \exp (-2\lambda t),
\end{equation}
with $\lambda\ge 0.$
A flow is called chaotic if 
$$
\lambda^\infty\equiv\lim_{t\rightarrow\infty}\lambda>0,
$$ 
otherwise it is said to be integrable.
We note that the laminar to turbulent transition is marked by the emergence
of many scales for the velocity field.
The integrable to non-integrable transition concerns the 
behavior of the fluid trajectories, not the spectrum of the
velocity field. This is why the stochasticity of a chaotic flow is also called
Lagrangian turbulence (Lagrangian trajectories), in analog with the usual 
Kolmogorov-type Eulerian turbulence (Eulerian flow field).
As far as the theory of advection-diffusion equation concerns,
a qualitative change in the solution of the
equation occurs at the integrable-nonintegrable 
transition. This should be contrasted with the standard
treatment that presumes a turbulent background flow.        

What we mean by a qualitative change in the solution of the 
advection-diffusion equation can be made precise
by the rate at which the gradients of
passive scalar field grow before the diffusive relaxation
dominates the solution. In an integrable flow, $\nabla\phi$
grows linearly in time. If the Lagrangian trajectories become
chaotic, $\nabla\phi$ increases exponentially in time.
The exponential rate is given by the finite time or local
Lyapunov exponent $\lambda(\xi,t),$ which depends on both position
and time. That is to say, the value of $\lambda$ depends on where the 
fiducial trajectory initially starts and how long one is tracing the 
fiducial trajectory. The time and especially the spatial dependence
of the finite time Lyapunov exponent define the essential 
characteristics of the passive scalar transport.

The effects of advection and diffusion can be most clearly seen
in Lagrangian coordinates. The Lagrangian coordinates are defined
by a one-to-one mapping between the initial position and the 
current position of a fluid element, which is found by
integrating 
$$
d{\x}(\xi,t)/dt = {\v}({\x},t)
$$
with the initial condition ${\x}(\xi,t=0)=\xi.$
The description of a physical phenomena is independent of the
choice of the coordinate system.     
Both ${\x}$ and ${\xi}$ can provide the necessary coordinate system,
the first is the usual Eulerian coordinates while the second 
one is the well-known Lagrangian coordinates.
Unlike the usual Eulerian coordinates, the Lagrangian coordinates
have a non-trivial metric tensor. In fact it has both a space and a time
dependence.
If the metric tensor of the Eulerian frame is the unit matrix,
the metric tensor of the Lagrangian coordinates is defined as
\begin{equation}
\label{metric_tensor_definition}
g_{ij} \equiv \partial{\x}/\partial\xi^i\cdot\partial{\x}/\partial\xi^j
\,\,\,\,
{\rm and}\,\,\,\,
g^{ij} \equiv \nabla\xi^i\cdot\nabla\xi^j.
\end{equation}
The metric tensor, by its definition, is entirely
determined by the flow field. Even for a perfectly smooth steady 
flow field, the metric tensor tends to become singular as $t$ becomes
large. If the flow is integrable, the metric tensor diverges at most
quadratically in time \cite{tang_pof}. 
For a chaotic flow, the metric tensor blows up exponentially
in some subspaces. As long as the flow field is well-behaved, the metric
tensor is well behaved and does not possess any finite time singularity.

Since the Lagrangian coordinates are attached to the fluid elements,
the solution to the ideal advection equation is equivalent to 
integrating the trajectory ${\x}(\xi,t)$ and setting 
$\phi({\x}(\xi,t),t) = \phi(\xi,t=0).$
If the spatial gradient of $\phi$ is written in Lagrangian coordinates,
$$
[\nabla\phi({\x},t)]^2 = \nabla_0\phi(\xi)\cdot\tensor{g}
\cdot\nabla_0\phi(\xi),
$$
the coordinate derivative in Lagrangian coordinates of $\phi,$
$\nabla_0\phi(\xi),$ would be time independent, {\it i.e.} given by initial
condition, $\nabla\phi({\x},t=0).$
When diffusion is included, $\phi(\xi,t)$ would have a time dependence,
so does $\nabla_0\phi(\xi,t).$
Two important characteristics can be immediately explained once
the covariant representation of the metric tensor, $g_{ij},$ 
is identified as the Oseledec matrix \cite{ruelle_review}.
Since $g_{ij}$ is a positive, definite symmetric matrix, 
it can be diagonalized with real eigenvectors and positive eigenvalues,
\begin{equation}
\label{metric_diagonal}
g_{ij} = \exp(2\lambda t) {\e}{\e} + \exp(-2\lambda t) {\s}{\s}
\end{equation}
with $\lambda\ge 0.$
The gradient of $\phi$ can be written in the form,
\begin{equation}
\label{gradient}
(\nabla\phi)^2 = (\nabla_0\phi\cdot {\e})^2 \exp(-2\lambda t)
	+ (\nabla_0\phi\cdot {\s})^2 \exp(2\lambda t).
\end{equation}
The metric tensor that enters the last calculation
is its contravariant form $g^{ij},$ which is the exact matrix
inverse of $g_{ij},$ hence 
$g^{ij} = \exp(-2\lambda t) {\e}{\e} + \exp(2\lambda t) {\s}{\s}.$
 
What immediately can be seen is that under ideal advection,
the passive scalar gradient would blow up exponentially in time
without bound since $\nabla_0\phi(\xi)$ is time-independent.
This reflects the singular nature of the diffusion term as a
perturbation to the ideal advection equation.
The existence of a diffusion term, no matter how small 
$D$ is, would eventually remove any gradient in $\phi.$
Since the metric tensor has an exponential factor, the diffusion
responsible for the relaxation of $\nabla_0\phi(\xi,t)$ must have
a super-exponential dependence in order to overcome the exponential
factor in the metric tensor.
The second note is that the diffusion is essentially one dimensional
since the exponential growing term in the metric tensor is only 
in the one dimensional subspace defined by ${\s}.$

\section{Time-dependent solution to the advection-diffusion equation}
\label{sec:closed_flow}

In this section 
we give an example for a bounded flow, {\it i.e.} a closed system.
The open flow configuration, the main focus of this paper, is treated 
afterwards. A closed system corresponds to an initial value problem.
The equation to solve is the advection-diffusion equation,
$$
\partial\varphi/\partial t + {\v}\cdot\nabla{\varphi}=
\nabla\cdot(D\nabla\varphi).
$$
We have defined
$$
\varphi\equiv \phi-\bar{\phi},\,\,\,\,{\rm and}\,\,\,\,\,
\bar{\phi}=\int \phi d^3{\x}.
$$
Transforming into Lagrangian coordinates,
$$
(\partial\varphi/\partial t)_\xi =
\nabla_0\cdot(D\tensor{g}\nabla_0\varphi).
$$  
The tensor diffusivity $\tensor{D}\equiv D\tensor{g}$ in diagonal
form is
\begin{equation}
\label{tensor_diffusivity_diagonal}
\tensor{D} = De^{2\lambda t} {\s}{\s} + De^{-2\lambda t} {\e}{\e}.
\end{equation}
We have restricted the discussion to a two dimensional flow.
The three dimensional case is treated in \cite{tang_pof}. 

The exponential anisotropy of the tensor diffusivity along different
directions implies that the diffusive relaxation is determined by
a one dimensional diffusion equation,
\begin{equation}
\label{diffusion-1d-t}
\partial\varphi/\partial t = \nabla_0\cdot D e^{2\lambda t} {\s}{\s}
\cdot \nabla_0\varphi.
\end{equation}
The usefulness of above formulation rests on the fact that
${\s}$ converges exponentially to a time asymptotic limit
${\s}_\infty(\xi).$ A time-independent vector field
${\s}_\infty(\xi)$ allows the usual construction of a
coordinate system that makes the calculation well-defined.
Parameterizing distance along the ${\s}$ line by a scalar field
$\beta,$ one has
\begin{equation}
\label{diffusion-1d-t-simple}
{\partial\varphi\over{\partial t}} = 
{\partial\over{\partial \beta}} D e^{2\lambda t} 
{\partial\varphi\over{\partial \beta}}.
\end{equation}
The exponential convergence rate of ${\s}$ to ${\s}_\infty$ 
implies that the calculation
based on the time-asymptotic spatial coordinates is exponentially 
accurate.

The solution to the one dimensional diffusion equation is determined 
by two quantities. One is the fundamental time scale of the problem,
the local Lyapunov time of the flow $1/\lambda.$
The other is the dimensionless number $\Omega\equiv \lambda L^2/D,$
which is the ratio between the characteristic diffusion time scale
$L^2/D$ and the Lyapunov time of the flow.
If $\Omega\gg 1$ which is generally true due to the smallness of the
molecular diffusivity, the scalar field undergoes a pure advection
until time $t_a\equiv (\ln2\Omega)/2\lambda.$
The ideal advection increases the gradients of the passive scalar
field (in Eulerian frame) by a factor of $\Omega.$
There is a rapid diffusive relaxation [super-exponential in 
$\partial\varphi(\xi,t)/\partial\beta$] which removes the gradients
$\nabla\varphi({\x},t)$ during a short interval of a few Lyapunov time
after $t_a.$

A clarification should be made on the definition of the Lyapunov 
time for passive scalar transport studies.
To our knowledge, in the literature (for a review, see \cite{ottino_review}), 
the Lyapunov time of a flow is uniformly 
associated with the stretch rate, {\it i.e.} 
the largest positive Lyapunov exponent of the flow.
This is a misconception and the reason can be easily seen by examining
equations (\ref{separation},\ref{metric_diagonal},\ref{gradient},
\ref{tensor_diffusivity_diagonal},\ref{diffusion-1d-t}).
The correct statement is that the most negative Lyapunov exponent $-\lambda$,
{\it i.e.} the convergence rate, defines the Lyapunov time $1/\lambda$ for
the passive scalar transport.  
In the case of a two dimensional time-dependent divergence-free flow which
happens to be
the starting point of much of the existing literature, the positive 
Lyapunov exponent has the same magnitude 
as the negative one due to the constraint of area-preserving, so this 
distinction is not mathematically important despite its physical significance.
However, in the more general case of a three dimensional flow or compressible
flows, a correct understanding of this subtlety is necessary both 
physically and mathematically. 

If the finite time Lyapunov exponent does not vary in space, 
then one has the ideal situation that the diffusive relaxation 
uniformly removes the gradients.
Unfortunately the only flow that is known to have this property
is purely hyperbolic. An example is the Arnold's cat map. 
Generic flows are non-hyperbolic and the finite time Lyapunov  
exponent has a peculiar spatial dependence.
Roughly speaking, $\lambda(\xi,t)$ varies smoothly along the ${\s}$ lines,
which is required for equations 
(\ref{diffusion-1d-t},\ref{diffusion-1d-t-simple}) to be well-posed.
The variation of $\lambda(\xi,t)$ across the ${\s}$ lines is pathological:
the gradient of $\lambda(\xi,t)$ in directions other than  
${\s}_\infty$  has an exponential dependence in time.
The smoothness of $\lambda$ along the ${\s}$ lines permits the existence
of a class of diffusion barrier in a chaotic region. The
pathology of $\lambda$ across the ${\s}_\infty$ direction gives rise
to the fractal-like transport in both space and time.

A crude estimate of the mixing process can be based on the probability 
distribution function of the finite time Lyapunov exponent.
If the distribution function is an $\delta$ function centered at
$\lambda^\infty.$ The relaxation of $(\nabla\phi)^2$ follows a simple route. 
The gradient grows by a factor of $\Omega$ and is then
removed during a Lyapunov time centered on time 
$t_a\equiv(\ln 2\Omega)/2\lambda.$
Equivalently the global mixing index $\sigma(t)$ 
stays the same until time $t_a,$ 
after which it exponentially decays to zero within a few Lyapunov time.
If the finite time Lyapunov exponent has a broadened distribution like
that shown in \cite{tang}, the time evolution of the global 
mixing index can be found 
by convoluting the ideal mixing curve with the $\lambda$ distribution function.
For a flow having noticeable amount of remnant integrable region,
there is usually a significant bump at the left end of the distribution
function. Since the distribution function is sampled over the entire space,
there are actually two types of trajectories which could 
contribute to this bump. The first is simply the integrable trajectory lying
on the KAM surfaces. The second one forms the so-called stochastic layer,
a sticky region surrounding the remnant KAM islands.
The bump in the distribution function 
 is responsible for the long tail of the $\sigma$ relaxation curve.
The experimental observation of a spread in $\sigma$ 
relaxation is determined by the $\lambda$ profile.

Although the features of the global solution is roughly given by the 
mean Lyapunov time, the $\Omega$ number associated with the local
Lyapunov time dictates the local details of the solution.
For example, since diffusion occurs only along an ${\s}$ line, the
place with a peculiarly small finite time Lyapunov exponent would 
pose as a practical diffusion barrier. 
A more precise description is based on the theory of finite time Lyapunov
exponent \cite{tang,tang_pla}.
The finite time Lyapunov exponent exponentially converges to a 
form of three parts:
\begin{equation}
\label{exponent_decomposition}
\lambda(\xi,t) = {\tilde{\lambda}\over{t}} + {f\over\sqrt{t}}
+ \lambda^\infty.
\end{equation}
The two convergence functions have entirely different properties.
The function $\tilde{\lambda}$ is a smooth function of space
and it is related to the geometry of the ${\s}$ line by
\begin{equation}
\label{s-exponent}
{\s}_\infty(\xi)\cdot\nabla_0\tilde{\lambda} + \nabla_0\cdot{\s}_\infty=0.
\end{equation}
The function $f(\xi,t)$ reflects the pathology in the spatial variation
of the finite time Lyapunov exponent.
First it does not vary along the ${\s}$ direction,
\begin{equation}
\label{exponent_fractal_s}
{\s}_\infty\cdot\nabla_0 f(\xi,t) = 0.
\end{equation}
Secondly it develops an exponentially growing gradients in the 
direction across the ${\s}_\infty,$ for example,
$$
{\e}_\infty\cdot\nabla_0 f \sim \exp(\lambda t).
$$
Of course, the time dependence of $f(\xi,t)$ has to be bounded
by $\sqrt{t}$ so $\lim_{t\rightarrow\infty} \lambda = \lambda^\infty.$

Although equation (\ref{s-exponent}) 
states that the finite time Lyapunov exponent
achieves an extreme when $\nabla\cdot{\s}_\infty$ vanishes, 
numerical calculations show that $\lambda$ reaches its local minimum 
where the ${\s}$ line makes sharp bend.  
The sharp drop in the magnitude of $\lambda$ implies that
these sharp bends of an ${\s}$ line are practical diffusion barriers.
We note that an ${\s}$ line belongs to a chaotic region, and they persist
even if the flow is far from integrable.
Equivalently speaking, even if the flow is driven globally chaotic without 
discernible KAM islands, there are still regions with retarded mixing.
Since the production of new chemical C depends on the final diffusive 
relaxation of $A$ and $B,$ these diffusion barriers directly affects
the quality of a reactor.

\section{steady state mixer}
\label{sec:steady_mixer}

An efficient engineering device employs an open flow operating at steady state.
The simplest geometry for such a device is a 
pipe \cite{chang_96,dean_27,loc_87} so we will
consider the cylindrical geometry.
The flow field is assumed to be divergence-free and quasi-two-dimensional,
\begin{equation}
\label{2.5flow}
{\v}(x,y,z) = {\zhat}\times\nabla\psi(x,y,z) + v_z(x,y){\zhat}.
\end{equation}
The absence of a $z$ dependence in $v_z(x,y)$ implies that the flow 
field is divergence-free in the transverse plane,
$\nabla_{xy} \cdot {\v}_{xy}(x,y,z)=0$ with 
${\v}_{xy}={\zhat}\times\nabla\psi(x,y,z).$
If the mixer is operated at steady state ($\partial\phi/\partial t=0$),
the governing equation is
\begin{equation}
\label{steady-equation}
v_z{\partial\phi\over{\partial z}} + {\v}_{xy}\cdot\nabla_{xy}\phi
= \nabla_{xy}\cdot D\nabla_{xy}\phi + {\partial\over{\partial z}}
D {\partial\over{\partial z}} \phi.
\end{equation}
We have separated the divergence of the diffusive flux into longitudinal
and transversal components. For an operating reactor, the longitudinal 
component becomes exponentially smaller than the transversal component
going downstream before the onset of diffusive pulse that removes the
transversal gradients.
In the usual case that $D$ is small and $R^2 v_z/LD\gg 1,$
the longitudinal diffusion term is negligible so
\begin{equation}
\label{steady-equation-simplified}
v_z{\partial\phi\over{\partial z}} + {\v}_{xy}\cdot\nabla_{xy}\phi
= \nabla_{xy}\cdot D\nabla_{xy}\phi.
\end{equation}
Here $R$ is the radius of the cylinder and $L$ is the total longitudinal
length of the device.

The in-fluxes of reactants $A$ and $B$ are
$$
{\cal F}_{A,B} \equiv \int v_z(x,y,z=0) C_{A,B}(x,y,z=0)dxdy.
$$
The flux for each chemical can be defined for any cross section in the reactor,
but only their difference  is important.
The differential flux of the reactants 
$$
F(z) = \int v_z(x,y)\phi(x,y,z) dxdy = \int v_z(x,y)(C_A-C_B) dxdy
$$
is a constant in the longitudinal direction,
$$
d F(z)/dz = \int v_z{\partial\phi(x,y,z)\over{\partial z}} dxdy
=0.
$$
For practical applications, an initial input is proper if
$F(z=0)=0.$ 
The degree of mixing at the transversal plane is given by
\begin{equation}
\label{transversal_mixing_measure}
\sigma(z) = {1\over 2}\int v_z(x,y)\phi^2(x,y,z) dxdy.
\end{equation}
This spatially dependent mixing index $\sigma(z)$ should be distinguished
from the global mixing index $\sigma(t)$ defined in equation
(\ref{global_mixing_index}). 
The reactants become better mixed as they go downstream. 
The rate of mixing is determined by the diffusive effect,
\begin{equation}
\label{mixing_rate_z}
{d\sigma\over{dz}} = -\int D\nabla_{xy}\phi\cdot\nabla_{xy}\phi dxdy.
\end{equation}
The percentage of the chemicals failed to react is roughly measured by
$$
{\cal R} = {\sigma(z=L)\over{\sigma(z=0)}}.
$$
The quality of the mixer(reactor) is measured by the $Q$-factor 
$$
Q \equiv {1\over {\cal R}} = {\sigma(z=0)\over{\sigma(z=L)}}.
$$
In the impractical limit of $L\rightarrow\infty$
the chemicals would react completely for a proper initial
input $F(z=0)=0$ and the $Q$-factor is infinite.
For a finite $L,$ the $Q$-factor is generally finite as well.
For an intake flux
$$
{\cal F}_A = \int v_z(z=0) C_A dxdy,
$$
the amount that does not react is about
$$
{\cal F}_A/\sqrt{Q}.
$$  
If $Q$ is infinite, the output of the final product is
$$
O_C = \int v_z(x,y,z=L) C_C(x,y,z=L) dxdy = {\cal F}_A.
$$
Otherwise
$$
O_C \approx (1-Q^{-1/2}) {\cal F}_A.
$$

The optimization of a mixing device or a reactor is primarily looking for
a balance between the production rate ${\cal F}_A,$ reactor length $L,$ and
the desired $Q$-factor.
Implicitly through the $Q$-factor there is also an energy consumption
penalty (power $P$) for driving a flow field ${\v}({\x}).$
These correspond to four design constraints:
1] geometrical constraint, i.e. the physical size limit for the device;
2] production rate constraint;
3] production quality constraint;
4] energy constraint. 
The first three are kinematic constraints since they are uniquely
determined by the flow field [Except that in a reactor, there is a minimal
length constraint due to the finite reaction time, {\it i.e.}
the system size has to be greater than $v_z/\kappa\langle f/2\rangle.$
This point was made earlier in the introduction].
The last one is a dynamical constraint and it requires solving the 
Navier-Stokes equation to relate the driving term to the flow field.

This paper addresses the kinematic constraints. The dynamical constraint
for the flow field is more case-dependent.
Nevertheless, a general statement can be made
 that it is usually less energy consuming
to produce a smooth, non-turbulent, but chaotic flow, than a turbulent one.
In the case of a tubular device, this is reflected by a small increase
in the pressure drop over the tube, or equivalently the pumping power,
 to induce a chaotic flow with satisfactory mixing properties \cite{acharya_92}.

The physical meaning of the kinematic constraints will become clear
once the solution to equation (\ref{steady-equation}) is found.
The most important property of the flow field in determining the kinematic 
constraint is the so-called local Lyapunov exponent, just like the
time-dependent initial value problem.
Instead of time, the $z$ coordinate will be used to parameterize a fluid
trajectory. The only constraint for $z$ to be a good scalar for parameterization
is that $v_z$ nowhere vanishes. This is naturally satisfied in our choice
of a quasi-two-dimensional flow field, $\partial v_z/\partial z =0.$
Any point satisfying $v_z(x_0,y_0)=0$ would imply that $v_z$ vanishes
along a straight line from $(x_0,y_0,z=0)$ to $(x_0,y_0,z=L),$
which strictly prohibits transport.
Hence $v_z$ would be made nonzero in a working device.
This is easily achievable by an infinitesimal longitudinal perturbation.

The singular nature of the solution to the 
equation (\ref{steady-equation-simplified}) can be
seen by a coordinate transformation.
The fluid trajectory is now parameterized by $z$ and given
by 
\begin{equation}
\label{flow_field_in_z}
{d x\over{d z}} = {v_x(x,y,z)\over{v_z(x,y)}};\,\,\,\,
{d y\over{d z}} = {v_y(x,y,z)\over{v_z(x,y)}}.
\end{equation}
For an arbitrary initial position $(x_0,y_0)$ at $z=0$ plane,
a trajectory of the fluid element can be traced to $(x,y)$ at $z$-plane.
The functional relationship between $x(x_0,y_0,z)$ and $y(x_0,y_0,z)$ is found
by integrating equation (\ref{flow_field_in_z}) with 
initial position $(x_0,y_0)$ from $z=0$ to $z.$

If the mixing equation is transformed into the $(x_0,y_0)$ coordinates,
one has
\begin{equation}
\label{steady_mixing_lagrangian}
({\partial\phi\over{\partial z}})_{x_0,y_0}
= {1\over{v_z(x(x_0,y_0,z),y(x_0,y_0,z))}} \nabla_{x_0y_0}\cdot 
D\tensor{g}\nabla_{x_0y_0} \phi(x_0,y_0,z).
\end{equation}
The metric tensor of the $(x_0,y_0)$ coordinate,$\tensor{g},$
has two forms, covariant and contravariant.
The contravariant form is defined as

\[
g^{ij} = \left(
\begin{array}{cc}
\nabla x_0\cdot\nabla x_0 & \nabla x_0\cdot\nabla y_0\\
\nabla x_0\cdot \nabla y_0 & \nabla y_0\cdot\nabla y_0 
\end{array}
\right)
\]

The covariant representation is given by
\[
g_{ij} = \left(
\begin{array}{cc}
{{\partial (x,y)}\over{\partial x_0}}\cdot{\partial (x,y)\over{\partial x_0}} & 
{\partial (x,y)\over{\partial x_0}}\cdot{\partial (x,y)\over{\partial y_0}}\\
{\partial (x,y)\over{\partial x_0}}\cdot{\partial (x,y)\over{\partial y_0}} & 
{\partial (x,y)\over{\partial y_0}}\cdot{\partial (x,y)\over{\partial y_0}}
\end{array}
\right)
\]

The metric tensor is a positive, definite matrix, so it can be diagonalized
with positive eigenvalues and real eigenvectors.
If the covariant form is written as
\begin{equation}
\label{metric_z_cov}
g_{ij} = \exp(2\eta z) {\e}{\e} + \exp(-2\eta z) {\s}{\s}
\end{equation}
with $\eta\ge 0,$
the contravariant form is
\begin{equation}
\label{metric_z_contra}
g^{ij} = \exp(-2\eta z) {\e}{\e} + \exp(2\eta z) {\s}{\s},
\end{equation}
since $g_{ij}$ and $g^{ij}$ are matrix inverse of each other.
The meaning of ${\e}$ and ${\s}$ can be made precise
by following the distance between neighboring two points,
$$
(dx)^2+ (dy)^2 = [dx_0,dy_0] g_{ij} [dx_0,dy_0]^T.  
$$
Hence two initial points separating along the ${\e}$ direction 
would diverge exponentially going downstream, but they would converge 
exponentially going downstream if their initial separation is along the 
${\s}$ direction.
Conventionally ${\e}$ is called the unstable direction and 
${\s}$ is the stable direction.
The function $\eta(x_0,y_0,z)$ which depends on the initial
position of the trajectory $(x_0,y_0)$ and the longitudinal 
ending point $z,$ is a local Lyapunov exponent measuring the 
exponential separation rate.
The inverse of $\eta(x_0,y_0,z)$ defines a local Lyapunov scale,
{\it i.e.} the longitudinal distance over which the separation of neighboring 
trajectories varies by one e-fold.    
This is also the longitudinal distance over which the gradient of a 
passive scalar field increases by one e-fold in the absence of diffusion.

The mixing-reaction process is easier to understand if
one transforms equation (\ref{mixing_rate_z}) into the $(x_0,y_0)$
coordinates, 
$$
{d\sigma(z)\over{dz}} = - \int D \nabla_{x_0y_0}\phi(x_0,y_0,z)
\cdot\tensor{g}\cdot
\nabla_{x_0y_0}\phi(x_0,y_0,z) J dx_0dy_0.
$$
The Jacobian of the $(x_0,y_0)$ coordinates is identically unity
if $v_{xy}$ is divergence-free.
Substituting the diagonal form of the metric tensor, 
equation (\ref{metric_z_contra}), into above equation, one obtains
\begin{eqnarray}
{d\sigma(z)\over{dz}} = - \int & D & \{[{\s}(x_0,y_0,z)
\cdot\nabla_{x_0y_0}\phi(x_0,y_0,z)]^2 \exp(2\eta z) \nonumber \\
& + & [{\e}(x_0,y_0,z)\cdot\nabla_{x_0y_0}\phi(x_0,y_0,z)]^2 
\exp(-2\eta z)\} dx_0dy_0.
\end{eqnarray}
The reduction in $\sigma(z)$ can be calculated using a simpler form
in the usual case $R^2v_z\eta/D\gg 1.$
The simplification comes from two factors: 1] 
the diffusive flux is negligible for small $z,$
2] ${\s}(x_0,y_0,z)$ exponentially loses its $z$ dependence, {\it i.e.}
${\s}(x_0,y_0,z)$ can be replaced by its $z$-asymptotic limit
${\s}_\infty(x_0,y_0)$ with an exponentially small correction.
Combining these, one can calculate the mixing rate with exponential
accuracy by
\begin{equation}
\label{steady_mixing_rate_final}
{d\sigma(z)\over{dz}} = - \int D [{\s}(x_0,y_0)\cdot\nabla_{x_0y_0}
\phi(x_0,y_0,z)]^2 \exp(2\eta z) dx_0dy_0.
\end{equation}
The gradient of $\phi(x_0,y_0,z)$ in the $(x_0,y_0)$ coordinates
is found by solving a simplified form of 
equation (\ref{steady_mixing_lagrangian})
\begin{equation}
\label{mixing_1d_z}
({\partial\phi\over{\partial z}})_{x_0,y_0}
= {1\over{v_z(x_0,y_0,z)}} 
{\partial\over{\partial\beta}}De^{2\eta(x_0,y_0,z)z}
{\partial\over{\partial\beta}} \phi(x_0,y_0,z).
\end{equation}
The $\beta$ coordinate is defined by a parameterization along the ${\s}$ lines
$$
d{\x}_0/d\beta \propto {\s}_\infty(x_0,y_0) \,\,\,\,{\rm with}\,\,\,\,{\x}_0=(x_0,y_0).
$$
The solution of this one dimensional equation differs from that of the 
full equation by an exponentially small term in $z.$
The discrepancy is negligible if $R^2 v_z\eta/D\gg 1$ and 
the initial gradient has a scale comparable to $R.$
  
The solution to equations (\ref{steady_mixing_rate_final},\ref{mixing_1d_z})
has remarkable properties for a chaotic flow ($\eta > 0.$)
Since the longitudinal distance plays the role of time in a steady state
system, the usual characteristic diffusion
time $R^2/D$ gives rise to an equivalent characteristic longitudinal length
for diffusion $R^2 v_z/D.$
The ratio of this characteristic longitudinal length for diffusion
and the Lyapunov length defines $\Omega,$
\begin{equation}
\label{Omega_def}
\Omega \equiv {R^2v_z\eta\over{D}},
\end{equation}
a dimensionless number that is typically much greater than one.
 
\section{Degree of mixing versus longitudinal length}
\label{sec:mixing_length}

Recall that the transversal mixing index $\sigma(z)$ is defined as
$$
\sigma(z) \equiv \int {1\over 2} \phi^2(x,y,z) dxdy.
$$
The variation of $\sigma$ downstream is given by
$$
{d\sigma(z)\over{dz}} 
= - \int D \nabla\phi(x,y,z)\cdot\nabla\phi(x,y,z) dxdy.
$$
$\sigma$ is a monotonically decreasing function which corresponds
to the time-irreversibility of the diffusion process.
Dividing above equation by $\sigma(0)$ leads to a 
dimensionless form   
$$
{1\over\sigma(0)}{d\sigma(z)\over{dz}}
= - {2\int D \nabla\phi(x,y,z)\cdot\nabla\phi(x,y,z) dxdy\over
	{\int v_z(x,y,z=0) \phi^2(x,y,z=0) dxdy}}.
$$
In the case that $\eta$ is a constant,
\begin{equation}
\label{mixing_versus_z}
{1\over\sigma(0)}{d\sigma(z)\over{dz}}
= - {D\over{v_z R^2}} \exp[-{{e^{2\eta z}-1}\over{2\Omega}} + 2\eta z].
\end{equation}

Equation (\ref{mixing_versus_z}) can be integrated for an exact analytic expression,
\begin{equation}
\label{mixing_index_ideal}
{\sigma(z)\over\sigma(0)} = 
\exp[-{{e^{2\eta z}-1}\over{2\Omega}}]
\end{equation}
with $\Omega$ defined in equation (\ref{Omega_def}).
This is a remarkable expression that is best interpreted in the units of
Lyapunov length.
There is little mixing and hence reaction in a region within a critical length
\begin{equation}
\label{minimum_length}
L_c = \ln(2\Omega)/2\eta
\end{equation}
from the intake of the device.
In fact, there is only one $e$-fold drop in $\sigma(z)/\sigma(0)$ over
the entire longitudinal length $L_c,$
$$
{\sigma(L_c)\over{\sigma(0)}} = e^{-1}.
$$
Each additional Lyapunov length beyond $L_c$ produces a super-exponential 
jump in the production quality,
e.g. at a length $L_c + N/\eta,$
$$
{\sigma(L_c+N\eta)\over{\sigma(0)}} =
\exp[{-e^N}],
$$
thus a reator of longitudinal length $L_c + N/\eta$ has a $Q$-factor
of
\begin{equation}
\label{ideal_Q}
Q = \exp[{e^N}].
\end{equation}
The above formula implies that it takes less than $N=3$ additional Lyapunov 
lengths to achieve a $Q$-factor of $10^8.$
For a reactor with a length of $L_c + 3/\eta,$ 
less than one part of ten thousands
input reactants fails to react at the moment of discharge.
It must be emphasized that $L_c$ is a modest number even if the reactants are
extremely difficult to mix, which is usually the case due to the smallness of
$D.$ For example, even if $\Omega=10^{10},\,L_c$ 
is about $12$ Lyapunov length $1/\eta.$
A reactor of $Q$-factor $10^{10}$ requires a total longitudinal 
length of $15$ Lyapunov lengths. 
In giving out these numbers, we have assumed that the chemicals undergo a
diffusion-limited reaction. Even if the reaction rate is so slow
that $L_c$ is less than a typical reaction length 
$v_z/\kappa\langle f/2 \rangle,$
an additional longitudinal length of a reaction length 
in the engineering design would provide the extra room to achieve
the desired $Q$-factor.   
   
The assumption that $\eta$ is a constant, is a deceptively nontrivial one.
In actuality it implies a much more stringent topological constraint.
It is known that only purely hyperbolic system can have a constant Lyapunov
exponent, the canonical example being the Arnold's cat map.
A point is hyperbolic if its stable and unstable directions are not degenerate.
Generic systems such as hamiltonian flows, are not purely hyperbolic. 
In fact, we suspect that a differomorphism higher that $C^2$ would generally have
nonhyperbolic points. In the case of hamiltonian systems, remnant KAM tori are
one manifestation of the nonhyperbolicity. Points on the chaotic set can also
be nonhyperbolic. In hamiltonian systems, they are responsible for
the sharp bending of the ${\s}$ lines.
 
\section{Local Lyapunov exponent and the $Q$-factor}
\label{sec:local_exponent_Q}

The local Lyapunov exponent, or equivalently the local Lyapunov length,
has a profound role in determining the quality of a reactor.
In retrospect, the design of a quality reactor can be considered as an
optimization of the local Lyapunov exponents.
The simplest way to understand the effect of a 
spatially varying Lyapunov exponent
on mixing is through its probability distribution function $P(\eta,z=L)$ for a 
steady state reactor.
If $\eta$ is a global constant, $P(\eta,z=L)$ would, of course, be a delta
function. Nonhyperbolicity prevents this for a generic flow. Instead there is a 
large spread in the distribution function.
The mixing index, or the $Q$-factor, can be approximated by convoluting
the distribution function with equation (\ref{mixing_index_ideal}).
For a device of longitudinal length $L$ 
operating with reactants of characteristic
diffusion time scale $R^2/D,$
there is a critical Lyapunov exponent $\eta_c(z=L)$ given by the solution to
equation
\begin{equation}
\label{lambda_c}
L = {1\over{2\eta_c(L)}} \ln {2v_z R^2 \eta_c(L)\over{D}}.
\end{equation}
As long as 
$$
\eta_c(L) > \eta^T(L) \,\,\,\,{\rm with}\,\,\,\,
\eta^T(z=L)\equiv {D e\over{2v_z R^2}},
$$  
the proportion of the input reactants that fails to react is roughly
given by 
$$
\int_0^{\eta_c(L)} P(\eta,z=L) d\eta,
$$
assuming that $P(\eta,z=L)$ is normalized
$$
\int_0^\infty P(\eta,z=L) d\eta = 1.
$$
A crude estimate for the $Q$-factor of the reactor is then
\begin{equation}
\label{Q-prob}
Q = 1/\int_0^{\eta_c(L)} P(\eta,z=L) d\eta.
\end{equation}

The second critical Lyapunov length $\eta^T(z=L) \equiv De/{2v_z R^2}$ 
is determined by 
$$
{\partial L_c(\eta)\over {\partial \eta}}_{\eta=\eta^T(z=L)}=0.
$$
$L_c$ is a monotonically decreasing function in $\eta$ if $\eta>\eta^T,$
but a monotonically increasing function if $\eta< \eta^T.$
The longer the characteristic diffusion time $R^2/D,$ the less
likely that the second critical scale $\eta^T(z=L)$ plays a role.
In other words, unless $R^2/D$ is small enough to render the 
chaotic mixer unnecessary, one does not need to worry about the subtleties
associated with $\eta^T.$

In the language of hamiltonian mechanics, the integrable region or
the KAM surfaces, are absolute barriers to advective transport because no
trajectories could cross them, at least for systems with no Arnold diffusion.
It is also an effective barrier for diffusive transport in the
advection-diffusion theory. This is explained by the time dependence
of the largest eigenvalue of the metric tensor.
It was shown\cite{tang_pof} that the largest eigenvalue of 
the metric tensor in an
integrable region grows at most quadratically in time, with 
the prefactor given by the shear rate of the surface\cite{tang_pof}, {\it i.e.}
the derivative of the rotational transform normal to a KAM surface.
Contrasted with the exponential growth in a chaotic region,
the effective diffusivity in $(x_0,y_0)$ coordinates,
$Dg^{ij},$ is small and hence the diffusive relaxation occurs on
a much longer time scale.

The simplest design optimization is to reduce the size of the remnant 
integrable regions. Once resonant perturbations are present,
global stochasticity can be achieved by increasing the 
perturbation strength. This is remarkable since it separates the global 
stochasticity for the trajectories from the spectrum of the velocity
field. A turbulent flow field is usually associated with a broad spectrum of
the velocity field. A laminar, but chaotic, flow has well-behaved 
velocity field, which is reflected by a few isolated spectral peaks.
The optimization involves the adjustment of the relative strength 
of the peaks.
The reduction of the integrable regions can then be realized under 
the constraint of maintaining the laminar nature of the flow.
 
The signature of a significant amount of integrable region in 
the local Lyapunov exponent is a large 
bump at the near-zero end of the distribution function.
The local Lyapunov exponent is also well-defined using the metric tensor
in an integrable region, even though the asymptotic Lyapunov exponent
vanishes. This is trivially explained by the fact 
that a quadratic function can always 
be approximated by an exponential function locally. 

\section{Geometrical constraint on a practical diffusion barrier} 
\label{sec:optimization_s_line}
 
Even if the integrable regions are so small that they are all invisible 
to naked eyes, or in the extreme, of size smaller than
$\sqrt{DL/v_z},$ the local Lyapunov exponent still has a finite 
spread in the distribution function.
What makes the matter worse is that the distribution function tends
to preserve an asymmetry biased towards the small $\eta$ end.
This statement is based on equations 
(\ref{exponent_decomposition},\ref{s-exponent},\ref{exponent_fractal_s}) 
and the numerical observation
that $\eta$ always takes a local minimum, rather than a local maximum,
 where the ${\s}$ line makes a 
sharp bend. Since the local Lyapunov exponent varies little where 
the ${\s}$ line is straight, but makes a sharp dip when the ${\s}$ makes
a sharp bend, there is an overall bias towards small $\eta$ in the 
distribution function, even if only the chaotic trajectories are included
in the distribution function calculation.

Surprisingly the presence of a small local Lyapunov 
exponent does not always lead to 
a practical diffusion barrier. To be a practical
diffusion barrier, not only a small effective diffusivity is required,
but also the size of the structure. Roughly speaking, the diffusivity $D$ sets
the minimal size of a practical diffusion barrier, 
although the integrability of
the trajectories also plays an important role.
In an integrable region, the ${\s}$ lines closes on itself forming closed
KAM curves. The shear-induced fast diffusion is confined within the KAM surfaces
so the radial diffusivity is just $D.$
The minimal size requirement of a practical barrier is given by
\begin{equation}
\label{min_barrier_kam}
w=\sqrt{DL/v_z}.
\end{equation}
It is the mean spread of an initial $\delta$-distribution at the center 
of an integrable island at the time of discharge.

In a chaotic region, diffusion occurs only along the ${\s}$ direction, but the
${\s}$ lines can not close on themselves. A single ${\s}$ line indeed fills an 
entire ergodic component densely. The previous estimate for the integrable case,
equation (\ref{min_barrier_kam}), usually is too small.
The minimal size for the diffusion barrier is comparable to
$$
w^2=\int_0^L D\exp(2\eta z) v_z^{-1} dz.
$$
The exact integration requires an exact form for $\eta,$
and the results are much more difficult to interpret.
We will use an idealized simple form for $\eta$ to illustrate
some of subtleties. Noting the empirical result
that $\eta \propto -(\ln\kappa)/z$ \cite{tang} 
with $\kappa$ the curvature of the ${\s}$ line bend,
we prescribe
$$
\eta = \eta_c - (c_0\ln\kappa)/z,
$$
with the understanding that $\kappa$ is scaled by $1/R$ and hence 
dimensionless.
Combining with equation (\ref{lambda_c}), one finds
\begin{equation}
\label{min_barrier_chaotic}
w = R/\kappa^{c_0}.
\end{equation}
This expression is not valid for large enough $\kappa$ that 
$\eta=\eta_c - (c_0\ln\kappa)/z$ approaching zero.
That gives the lowest bound $w\sim\sqrt{D/v_z\eta_c},$ consistent
with the worst scenario predicted by equation (\ref{min_barrier_kam}).
Otherwise, equation (\ref{min_barrier_chaotic}) gives a minimal
barrier width estimate that depends on the sharpness of the bends.

The quantity that $w$ should be compared with, is the size of the region
having a significantly smaller local Lyapunov exponent.
This region 
can be estimated as follows\cite{tang}.
The position $(x_0^c,y_0^c)$ with a local minimum for $\eta$ is solved 
from
$$
\kappa_e(x_0^c,y_0^c) = 0 \,\,\,\,
{\rm and} \,\,\,\,
\nabla_{x_0y_0}\cdot\kappa_e(x_0^c,y_0^c) > 0,
$$
where $\kappa_e$ is the curvature of the ${\e}$ line
$$
\kappa_e \equiv {\e}_\infty\cdot\nabla_0{\e}_\infty = 
- (\nabla_{x_0y_0}\cdot{\s}_\infty) {\s}_\infty.
$$
The core of the diffusion barrier is approximately 
bounded by the curve satisfying
$$
\nabla_{x_0y_0}\cdot\kappa_e(x_0^b,y_0^b) + \kappa_e^2(x_0^b,y_0^b)=0.
$$
$(x_0^b,y_0^b)$ is the closest point from $(x_0^c,y_0^c)$ satisfying
the above constraint.
Since the diffusion barrier is a sharp bend of an ${\s}$ line,
the width of the diffusion barrier is the inverse of the curvature
at $(x_0^b,y_0^b),$ {\it i.e.} $1/\kappa_e(x_0^b,y_0^b).$
One conclusion can be drawn immediately. 
Those ``diffusion barriers'' with a size
\begin{equation}
\label{geometrical_constraint}
1/\kappa_e(x_0^b,y_0^b) < \sqrt{DL/v_z}
\end{equation}
are not practical diffusion barriers affecting the $Q$-factor of a reactor.
A relaxed criteria is obtained by comparing $1/\kappa_e(x_0^b,y_0^b)$
with $w$ in equation (\ref{min_barrier_chaotic}). This implies that the local
geometry of the bends prohibits a simple resolution and explicit 
case-dependent calculation is required.

The cut-off size for a practical diffusion barrier brings a natural
limit on the finest grid size in a numerical calculation.
It also opens the possibility for a paradoxical statement that a smaller
$\eta$ in the distribution function might not imply poorer mixing
because the geometrical constraint, 
equations  (\ref{min_barrier_kam},\ref{min_barrier_chaotic}).
Consequently equation (\ref{Q-prob}) might not be a proper estimate for 
a globally chaotic flow.

An appropriate formula for estimating the $Q$-factor is based on the 
exact spatial dependence of the local Lyapunov exponent, $\eta(x_0,y_0,L).$
The formula is based on the integral
$$
\int \Theta(\eta^c-\eta(x_0,y_0,z)) dx_0dy_0,
$$ 
where $\Theta(x)$ is the step function
$$
\Theta(x) = 0 \,\,\,{\rm for}\,\,\, x<0; 1\,\,\, {\rm for}\,\,\, x>0.
$$
The geometrical constraint, equation (\ref{geometrical_constraint}) or alike,
is incorporated by imposing a fixed grid size of $\sqrt{DL/v_z}$ 
for calculating the above integral.
Written in discrete form,
\begin{equation}
\label{Q-spatial}
Q = {Av_z\over{ DL\sum_i\sum_j \Pi(\eta^c-\eta(i,j,L))}}  
\end{equation}
where $A$ is the total area of the cross section and the summations 
are over all grid points on a cross section.
In the case that $\Omega\gg 1,$ $\sqrt{DL/v_z}$ can be tiny compared
with the transversal scale of the device $R.$
Equation (\ref{Q-spatial}) is the proper formula for calculating
the $Q$-factor in a highly chaotic flow.
It should also yield a better estimate for the near-integrable cases,
but might not be necessary.
In principle, if the flow field ${\v}(x,y,z)$ is known, one would
calculate $\eta(x_0,y_0,L)$ and use equation (\ref{Q-spatial}) for
$Q.$ The main usage of a probability distribution function based 
formula like equation (\ref{Q-prob}) is in experimental situations 
where an exact measurement of $\eta(x_0,y_0,L)$ is inaccessible. 
$P(\eta,z=L)$ can be easily approximated by a limited amount of measurements.
For those who could afford the time and resources,
all quantities can be found to exponential accuracy by solving
equations (\ref{steady_mixing_rate_final},\ref{mixing_1d_z})
for a given velocity field ${\v}(x,y,z).$

In retrospect, a fully chaotic reactor can be further optimized by
balancing the value of the local Lyapunov exponent and the sharpness 
of the ${\s}$-bend. The basic idea is to either straight out the ${\s}$
line or make the bends sufficiently sharp that equation
(\ref{min_barrier_chaotic}) is violated. 

\section{A specific numerical example}
\label{sec:num_example}

Since the goal of this paper is to demonstrate the basic 
principles rather than working with a specific device, the illustrative
example will be chosen as simple as possible and computationally as 
efficient as possible. 
Nevertheless, there are a few features that we do intend to include
to make it practically relevant.
First we want the example flow to have a bounded chaotic region since
it is supposed to be confined by a pipe.
The flow is also expected to be stochastic along the axis.
Divergence-free is another useful feature to be consistent with earlier
analysis.

The simplest chaotic flow of the form equation (\ref{2.5flow}) has
$v_z$ a constant and $\psi$ a periodic function of $z.$
The standard treatment of this class of flow is to construct a
mapping by sampling at the period of $\psi$ along $z$ axis.
The mapping thus generated preserves both the topology of the Lagrangian
trajectories, the functional form of the local Lyapunov exponents (except
$z$ takes discrete values), and
the geometry of the ${\s}$ lines. 
Mapping is enormously more efficient than the flow in computations.
Hence our example will employ a mapping directly, but with the understanding
that it was reduced from a constant-$v_z$ and $z$-periodic-$\psi$ 
smooth three dimensional flow.
 
It should be noted right away that in a practical application,
an analytic form of the flow field is unlikely available, 
let alone an exact reduction to a mapping.
However, this simplification is just for the convenience of
illustration, the lack 
of it does not prevent a practical calculation.

There are three equations mapping $(x_n,y_n,z_n)$ to 
$(x_{n+1},y_{n+1},z_{n+1}).$ One of them is simply
$$
z_{n+1} = z_{n} + \Delta z
$$
with $\Delta z$ the period of $\psi.$
To insure area conservation, the example mapping that 
relates $(x_n,y_n)$ to $(x_{n+1},y_{n+1})$ 
is defined by a generating function, which has one free parameter
$k,$   
$$
S(x_n,y_{n+1}) = x_n y_{n+1} + k \ln(1+x_n^2 + y_{n+1}^2).
$$
Here $k$ signifies the perturbation.
The map is given by
\begin{eqnarray}
x_{n+1} & = & {\partial S\over{\partial y_{n+1}}} = x_n + 
2k {y_{n+1}\over{1+x_n^2 + y_{n+1}^2}}; \label{map_x} \\
y_n & = & {\partial S\over{\partial x_n}} = y_{n+1} + 
2k{x_n\over{1+x_n^2 + y_{n+1}^2}}. \label{map_y}
\end{eqnarray}
For $k\ll 1$ this is the twist map with the angle of twist per iteration
or every longitudinal advance of $\Delta z,$ equal to $2k/(1+x^2+y^2).$
The map has a non-zero Lyapunov exponent near the axis,
$x_n^2 + y_{n+1}^2 \ll 1,$ if $k$ is greater than unity.

\begin{figure}
\centerline{\psfig{figure=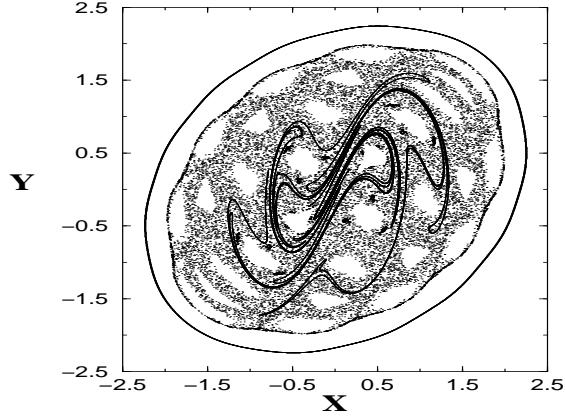,height=3.5in,width=3.5in}}
\caption{Poincare plot of the map given in equations (\ref{map_x},\ref{map_y})
	with $k=1.4.$ The solid line is an ${\s}$ line on $(x_0,y_0)$ plane.}
\label{figure:poincare_plot}
\end{figure}

This map is made explicit by solving a cubic equation for $y_{n+1}$
in terms of $x_n$ and $y_n.$ To do this, define   
$s$ so $y_{n+1}=s + y_n/3,$ which implies
$y_{n+1}^2=s^2+2s y_n/3 + y_n^2/9.$
One then finds
$$
s-{2\over 3}y_n + {2kx_n\over{1+x_n^2+{1\over 9}y_n^2+{2\over 3}sy_n
+ s^2}} = 0,
$$
which implies
$$
s^3 + s(1+x_n^2+{1\over 9}y_n^2) - {4\over 9}sy_n^2 
+ 2kx_n - {2\over 3}y_n(1+x_n^2+{1\over 9}y_n^2) =0.
$$
Define
$$
a=1+1+x_n^2-{1\over 3}y_n^2\,\,\,\,
{\rm and}\,\,\,\,
b = 2kx_n-{2\over 3}y_n(1+x_n^2+{1\over 9}y_n^2).
$$
The equation for $s$ can then be written in the standard form for a
cubic, $s^3+as + b = 0.$
This equation has an unique real root if
$$
r^2 \equiv ({b\over 2})^2 + ({a\over 3})^3
$$
is positive, the situation for this map. 
The real root of the cubic is then
$$
s = (r-{1\over 2})^{1\over 3} - (r+{1\over 2}b)^{1\over 3}.
$$

It is straightforward to show that the map is stochastic near the axis for 
$k>1.$ 
Near the axis, $x_n^2+y_{n+1}^2 << 1,$ the map
reduces to $y_{n+1}=y_n-2kx_n$ and $x_{n+1}=x_n + 2ky_{n+1}.$
This map is linear and can be solved by
$x_n = x_0\Gamma^n$ and $y_n=y_0\Gamma^n.$
One finds $(\Gamma-1)y_0=-2kx_0$ and $(\Gamma-1)x_0=2k\Gamma y_0$
or $(\Gamma-1)^2 + (2k)^2\Gamma=0.$ So
$$
\Gamma = (1-2k^2)\underline{+} 2\sqrt{k^2(k^2-1)}.
$$
Since the map is area preserving the two roots must satisfy 
$\Gamma_+\Gamma_-=1.$ If $k>1,$ one of the roots, $\Gamma_-,$
satisfies $\|\Gamma\|> 1$ and the map is stochastic
with a Lyapunov exponent $\ln\|\Gamma_-\|.$

Figure \ref{figure:poincare_plot} plots the intercepts of trajectories on the 
$(x_0,y_0)$ plane. The topology of this intercepts tells the 
integrability of the trajectories going downstream.
Since the flow field is periodic in $z,$ the cross sections
at $z=N\Delta z$ would have the same topology as that of $(x_0,y_0)$
plane. Because of this, an efficient way to understand the 
topology is by projecting the intercepts of a long trajectory
with cross sections at $z=N\Delta z, N=0,1,2,\cdots,$
on the $(x_0,y_0)$ plane.
Since we already have the proper mapping, this becomes simply plotting
the trajectories of the two dimensional mapping 
$(x_n,y_n)\rightarrow(x_{n+1},y_{n+1}).$
This is exactly how figure \ref{figure:poincare_plot} is generated.
Obviously our example map 
has a bounded chaotic region and it is stochastic near the
axis.

\begin{figure}
\centerline{\psfig{figure=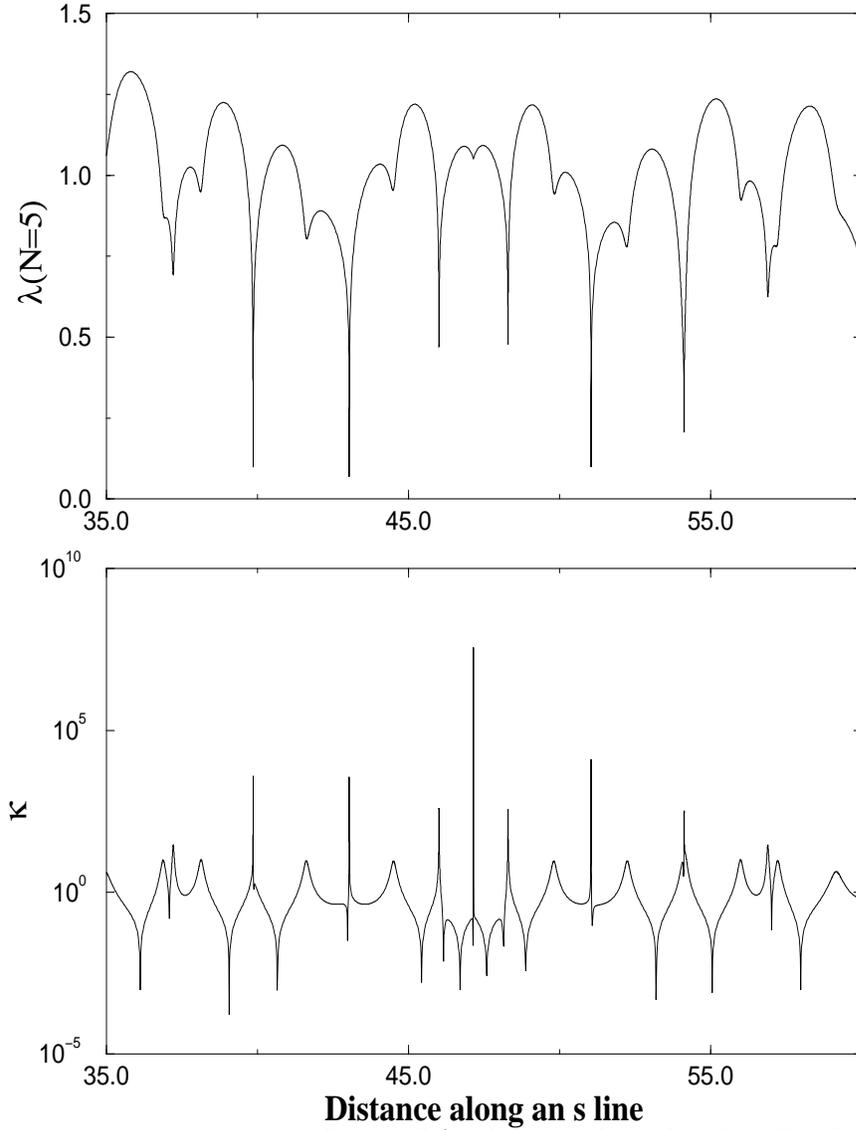,height=6.0in,width=3.5in}}
\caption{The local Lyapunov exponent at $z=N \Delta z$ with $N=5$ 
($\Delta z$ is the spatial period of the flow field in $z$ direction)
and the curvature $\kappa$
of the ${\s}$ line are plotted along an ${\s}$ line shown in 
figure \ref{figure:poincare_plot}. The local Lyapunov exponent makes a 
sharp dip wherever the ${\s}$ line makes a sharp turn. The missing dips
at extremely large curvature are due to finite machine precision 
and integration stepsizes. 
}
\label{figure:exponent_s_line}
\end{figure}

The most important feature we wish to demonstrate here is the 
geometry of the ${\s}$ lines and the spatial variation of the 
local Lyapunov exponent.
The ${\s}(x_0,y_0)$ line sits on the $z=0$ cross section.
The statement that diffusive relaxation only occurs along  
the ${\s}$ lines in $(x_0,y_0)$ coordinates is equivalent to say that,
initial chemical reactants located along an ${\s}$ line will come close
together and react to form new product $C_C.$
The sharp bends of the ${\s}$ line give rise to a peculiarly small
local Lyapunov exponent which hinders diffusive relaxation.
This relationship between the local Lyapunov exponent and the 
${\s}$-bends is shown in Figure \ref{figure:exponent_s_line}.
Wherever the curvature of the ${\s}$ goes up sharply, there is a 
significant dip in the local Lyapunov exponent.       

\section{Conclusions}
\label{sec:conclusion}

We have considered the design criteria of a chemical reactor device 
based on a chaotic flow. Particular attention has been paid to
a steady state reactor, where the Lyapunov length provides 
the fundamental spatial scale for the problem. 
The advantages of a chaotic flow for a steady-state reactor 
can be summarized into two scaling relations:
1] the minimum longitudinal length of a
reactor, $L_c,$ has a logarithmic dependence on the diffusivity $D,$
equation (\ref{minimum_length});
2] the quality of the reactor, the so-called $Q$-factor,
has a super-exponential dependence on each additional Lyapunov length
beyond $L_c,$ equation (\ref{ideal_Q}). 

The Lyapunov length $1/\eta,$ 
as defined through equations (\ref{flow_field_in_z},
\ref{metric_z_cov}), is a local Lyapunov length  
since it depends on the initial position $(x_0,y_0)$ and the longitudinal
displacement $z$ of the test fluid point.
The local Lyapunov length are defined everywhere in the $(x_0,y_0)$ plane,
independent of whether the trajectory is chaotic or integrable.
If the global flow field is known, the $Q$-factor of the reactor
can be calculated via equation (\ref{Q-spatial}).
In the experimental situation that the global flow field is not available, 
one can approximate the probability distribution function of the local
Lyapunov exponent (length) by repeated measurements over time.
The $Q$-factor can then be estimated by equation (\ref{Q-prob}). 
Although the derivation is presented with the assumption 
that flow field obeys equation (\ref{2.5flow}), these formulae
are expected to be useful under more general conditions, such as 
a weak $z$-dependence of $v_z$ and a weak time dependence of ${\v}.$
The fast reaction scenario, equation (\ref{fast_reaction}),
which brought mathematical simplicity to the discussion, can also be relaxed.    
A naive way to approach this is assuming that there is no reaction till
length $L_c+N/\eta,$ after which the reaction is suddenly turned on.
If the fast reaction condition is satisfied, near-perfectly mixed chemicals
$A$ and $B$ immediately react to form $C.$
Otherwise there is a delay to achieve the same $Q$-factor. The additional
length for the reactor is the  reaction length, the product of
the reaction time $1/\kappa\langle f/2 \rangle$ 
and the longitudinal flow velocity $v_z.$ 

For a closed flow reactor such as a stirred tank, the results obtained for
the steady-state device can be straightforwardly translated by identifying
the longitudinal length $z$ as time $t.$
The time scales replace the spatial scales as the quantities of
concern. For example,
the Lyapunov length $1/\eta$ is replaced by the Lyapunov time $1/\lambda$ 
and the crtical length $L_c$
is replaced by a critical advection time $t_a.$
One difference is that in a steady-state tubular device, the streamwise
distance can always be used to reduce the problem to a two dimensional one,
equation (\ref{flow_field_in_z}), but the time-dependent closed flow reactor
usually has to deal with a truely three dimensional flow, in addition to
a possible time dependence. Fortunately, aside from some additional subtleties,
the main physical charatersitics of the transport of 
a passive scalar in a three dimensional flow 
is the same as the two dimensional case \cite{tang_pof}.

\acknowledgements

We would like to thank U. S. Department of Energy for support under grant
DE-FG02-97ER54441. Part of the paper was written while one of the authors
(Tang) was supported by a NSF University-Industry Postdoctoral Fellowship
in Mathematical Sciences through SUNY Stony Brook.

\end{document}